\documentclass[twocolumn,twoside,slac_two]{revtex4}
\usepackage{graphicx}
\usepackage{fancyhdr}
\newcommand{\btokpiog}{B^0 \to K_S^0 \pi^0 \gamma}
\newcommand{\btokrhog}{B^0 \to K_S^0 \rho^0 \gamma}
\newcommand{\btopkog}{B^0 \to \phi K^0 \gamma}
\newcommand{\btopksg}{B^0 \to \phi K_S^0 \gamma}
\newcommand{\btopkpg}{B^+ \to \phi K^+ \gamma}

\begin{document}
\title{Exclusive Radiative {\boldmath $B$} meson decays at Belle}

%
\author{Himansu Sahoo on behalf of the Belle Collaboration}
\affiliation{Department of Physics and Astronomy, University of Hawaii, Honolulu, HI 96822, USA\\
himansu@phys.hawaii.edu}
\begin{abstract}
In this proceeding, we discuss recent results on exclusive radiative
$B$ meson decays from the Belle Collaboration. These decays are sensitive
to right-handed currents from New Physics. In particular, we measure
time-dependent $CP$ violation parameters in $\btokpiog$ and $\btokrhog$
decays, using high-statistics data samples collected  
at the $\Upsilon(4S)$ 
resonance with the Belle detector at the KEKB 
asymmetric-energy $e^+e^-$ collider. With the present statistics, 
these measurements are consistent with the standard model predictions. 
We also search for the radiative decay $\btopkog$
and report the first observation with a significance of $5.4\,\sigma$,
including systematic uncertainties. 
\end{abstract}

\maketitle

\thispagestyle{fancy}

\section{Introduction}
\begin{figure}[h]
\centering
\includegraphics[width=50mm]{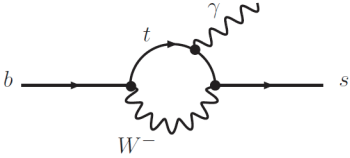}
\vspace{-0.1cm}
\caption{
Feynman diagram for the radiative $b\to s\gamma$ decays, 
showing the SM loop process with the $t$-quark contribution. 
} 
\label{fig:fey}
\end{figure}

Rare radiative decays of B mesons play an important role in the 
search for physics 
beyond the standard model (SM) of electroweak interactions. 
These flavor changing neutral current decays are
forbidden at tree level in the SM, but allowed through the 
electroweak penguin processes as in Fig.~\ref{fig:fey}. 
Hence, they are sensitive to non-SM particles mediating the loop 
(for example, charged Higgs or SUSY particles), which could 
affect either the branching fraction or $CP$ violation.
\par In the SM, the photon emitted from a 
$B^0$ ($\overline{B}{}^0$) meson in the radiative $b\to s\gamma$ decays,
is predominantly right-handed (left-handed). Therefore, the 
polarization of the photon carries information on the original 
$b$ flavor and  the time-dependent $CP$ asymmetry is suppressed 
by the quark mass ratio ($2m_s/m_b$)~\cite{ags1}.
In several models beyond the SM, the photon acquires an appreciable 
right-handed component due to the exchange of a virtual heavy 
fermion in the loop process, 
resulting in large values of mixing-induced $CP$ asymmetries.
The same argument holds in any multibody final states 
$B^0 \to P^0 Q^0 \gamma$, where $P^0$ and $Q^0$ are 
charge-conjugate states~\cite{ags2} 
(e.g., $\btokrhog$, $\btopksg$)~\cite{conj}.
A non-zero value of $CP$ asymmetry will 
be a clear hint of new physics.
\section{Experimental Apparatus}
\par The Belle detector is a large-solid-angle magnetic
spectrometer that consists of a silicon vertex detector (SVD),
a 50-layer central drift chamber (CDC), an array of
aerogel threshold Cherenkov counters (ACC),
a barrel-like arrangement of time-of-flight
scintillation counters (TOF), and an electromagnetic calorimeter (ECL)
comprised of CsI(Tl) crystals located inside
a superconducting solenoid coil that provides a 1.5~T
magnetic field.  An iron flux-return located outside 
the coil is instrumented to detect $K_L^0$ mesons and to identify
muons (KLM).  The detector is described in detail elsewhere~\cite{Belle}.
Two different inner detector configurations were used. 
For the first sample
of $152 \times 10^6$ $B\overline{B}$ pairs, a 2.0 cm radius beampipe
and a 3-layer silicon vertex detector (SVD1) were used;
for the latter samples,
a 1.5 cm radius beampipe, a 4-layer silicon detector (SVD2),
and a small-cell inner drift chamber were used.
\section{Analysis Technique }
At the KEKB asymmetric-energy $e^+ e^-$ (3.5 on 8.0 GeV) 
collider~\cite{kekb}, the $\Upsilon(4S)$ is produced with a Lorentz 
boost of $\beta \gamma = 0.425$ nearly along the $z$ axis, which 
is defined as opposite to the $e^+$ beam direction.
In the decay chain 
$\Upsilon(4S)\to B^0 \overline{B}{}^0 \to f_{\rm rec} f_{\rm tag}$, 
where one of the $B$ mesons decays at time $t_{\rm rec}$ 
to a final state $f_{\rm rec}$, which
is our signal mode, and the other decays at time $t_{\rm tag}$ 
to a final state $f_{\rm tag}$ that distinguishes between 
$B^0$ and $\overline{B}{}^0$, the decay
rate has a time dependence given by
\begin{eqnarray}
{\cal P}(\Delta{t})= \frac{ e^{-|\Delta{t}|/{\tau_{B^0}}} }{4\tau_{B^0}}
\biggl\{1 & + & q \cdot 
 \Bigl[ {\cal S} \sin(\Delta m_d \Delta{t})  \nonumber \\
 & + & {\cal A} \cos(\Delta m_d \Delta{t})
\Bigr] \biggr\}.
\label{eq_decay}
\end{eqnarray}
\noindent Here $\mathcal{S}$ and $\mathcal{A}$ are the 
$CP$-violation parameters,
$\tau_{B^0}$ is the neutral $B$ lifetime,
$\Delta m_d$ is the mass difference 
between the two neutral $B$ mass eigenstates,
$\Delta t = t_{\rm{rec}} - t_{\rm{tag}}$,
and the $b$-flavor charge $q$ equals $+1$ ($-1$) 
when the tagging $B$ meson is identified as 
$B^0$ ($\overline{B}{}^0$).
Since the $B^0$ and $\overline{B}{}^0$ are approximately 
at rest in the $\Upsilon(4S)$ center-of-mass system (cms),
$\Delta t$ can be determined from the displacement in $z$ 
between the $f_{\rm rec}$ and $f_{\rm tag}$
decay vertices: $\Delta t \simeq \Delta z / (\beta \gamma c)$,
where $c$ is the speed of light.
\subsection{Flavor Tagging and Vertex Reconstruction}
The $b$ flavor of the accompanying $B$ meson is identified 
by a tagging algorithm~\cite{tag} that categorizes 
charged leptons, kaons, and $\Lambda$
baryons found in the event.
The algorithm returns two parameters: the 
$b$-flavor charge $q$, and $r$, which 
measures the tag quality
and varies from $r=0$ for no flavor discrimination to 
$r = 1$ for unambiguous flavor assignment.
If $r < 0.1$, the accompanying $B$ meson provides negligible
tagging information and we set the wrong tag probability to 0.5.
Events with $r > 0.1$ are divided into six $r$ intervals.
The wrong tag fractions for the six $r$ intervals,
$w_l$ ($l = 1, 6$) and possible differences in $w_l$ 
between $B^0$ and $\overline{B}{}^0$
decays ($\Delta w_l$) are determined using
high-statistics control samples of semi-leptonic
and hadronic $b\to c$ decays~\cite{belle_cc,belle_b2s}.
\par The vertex position of the signal-side decay
is reconstructed from the $K_S^0$ trajectory in $\btokpiog$
and from the daughters of the $\rho^0$ in $\btokrhog$ mode,
with a constraint on the interaction point. 
The tracks are required to have enough hits in the SVD for vertexing.
The tag-side $B$ vertex is determined from well reconstructed
tracks that are not assigned to the signal side.
\par We determine $\mathcal{S}$ and 
$\mathcal{A}$  by performing
an unbinned maximum-likelihood (UML) fit to the observed 
$\Delta t$ distribution.
The likelihood function is
\begin{eqnarray}
\mathcal{L}(\mathcal{S},\mathcal{A})=
\prod_{i}\mathcal{P}_i(\mathcal{S},\mathcal{A};\Delta t_{i}),
\label{likelicpfiteq}
\end{eqnarray}
where the product includes all events in the fit.
The probability density function (PDF) is given by
\begin{eqnarray}
{\cal P}_i
&=& (1-f_{\rm ol}) \int 
\biggl[
\sum_j f_j {\cal P}_j(\Delta t') R_j(\Delta t_i-\Delta t') 
\biggr]
d(\Delta t')
\nonumber \\
&+&f_{\rm ol} P_{\rm ol}(\Delta t_i).
\label{eq:likelihood}
\end{eqnarray}
where $j$ runs over the signal and all background components.
The fractions of each component ($f_j$) depend on the $r$ region 
and are calculated on an event-by-event basis as a function of the
fitted variable. $R_j$ is the $\Delta t$ resolution function and
$P_{\rm ol}(\Delta t)$ is a broad Gaussian function that 
represents an outlier component with a small fraction $f_{\rm ol}$. 
The only free parameters in the final fit are ${\cal S}$ and 
${\cal A}$, which are determined by maximizing the likelihood 
function given by Eq.~\ref{likelicpfiteq}.
We define the raw asymmetry in each $\Delta t$ bin by 
$(N_{+}-N_{-})/(N_{+}+N_{-})$, where $N_{+}$ $(N_{-})$
is the number of observed candidates with $q=+1$ $(-1)$.
\section{\label{kspi0g} Time-dependent Analysis of {\boldmath $\btokpiog$} }
This analysis is done in Belle using $535 \times 10^6$ 
$B\overline{B}$ pairs~\cite{ushiroda-kspi0gamma}.
Since the time-dependent $CP$ asymmetry
is not expected to change significantly as a
function of $K_S\pi^0$ invariant mass,
we perform two measurements:
one for $B^0 \to K^{*0}(\to K_S^0 \pi^0)\gamma$
by requiring $M_{K_S^0\pi^0}$ to
lie in the range 
$0.8 < M_{K_S^0\pi^0} < 1.0 \;{\rm GeV}/c^2$,
and the other for
the full range of $M_{K_S^0 \pi^0}$ below $1.8 \;{\rm GeV}/c^2$.
\par The primary signature of these type of decays is the 
high energy prompt photon.
They are selected from isolated ECL clusters,
with center-of-mass (cms) 
energy in the range $1.4$ to $3.4$ GeV.
The polar angle of the photon direction in the laboratory frame is
restricted
to the barrel region of the ECL ($33^\circ < \theta_\gamma <
128^\circ$) for SVD1 data, but is extended to
the end-cap regions ($17^\circ < \theta_\gamma < 150^\circ$)
for SVD2 data due to the reduced material in front of the ECL.
The selected photon candidates are required to be consistent with 
isolated electromagnetic showers, i.e., $95\%$ of the 
energy in an array of
$5 \times 5$ CsI(Tl) crystals should be concentrated in 
an array of $3 \times 3$ crystals and 
should have no charged tracks associated with it.
We also remove photons from $\pi^0$($\eta$) $\to \gamma \gamma$ 
using a likelihood function described in Ref.~\cite{pi0etaveto}.
Neutral kaons ($K_S^0$) are reconstructed from two 
oppositely charged pions
that have an invariant mass within $\pm 6\; {\rm MeV}/c^2$ of 
the $K_S^0$ mass. Neutral pions ($\pi^0$) are formed from 
two photons with an
invariant mass within $\pm 16\;{\rm MeV}/c^2$ of the $\pi^0$ mass.
\begin{figure}[h]
\centering
\includegraphics[width=40mm]{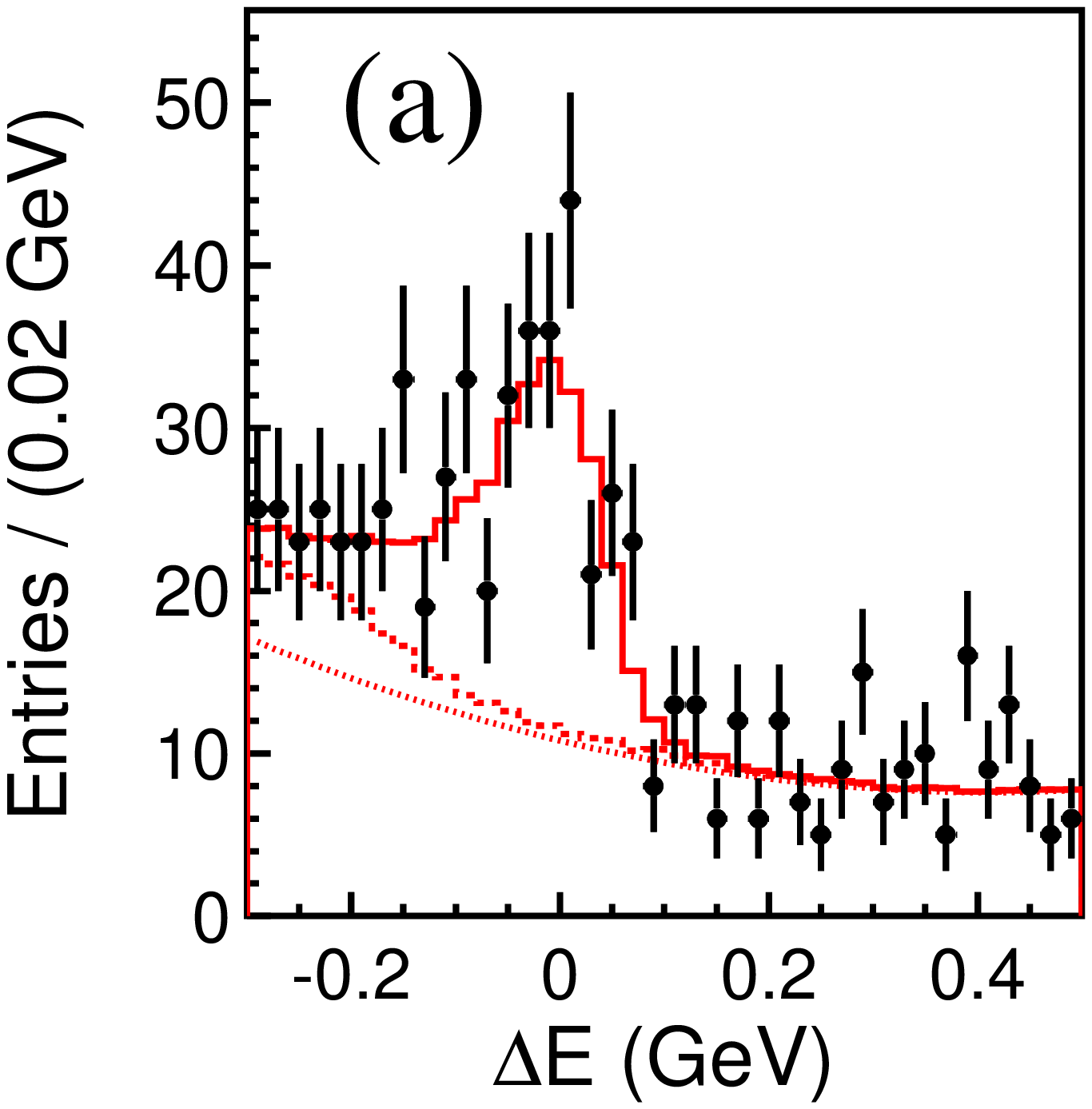}
\includegraphics[width=40mm]{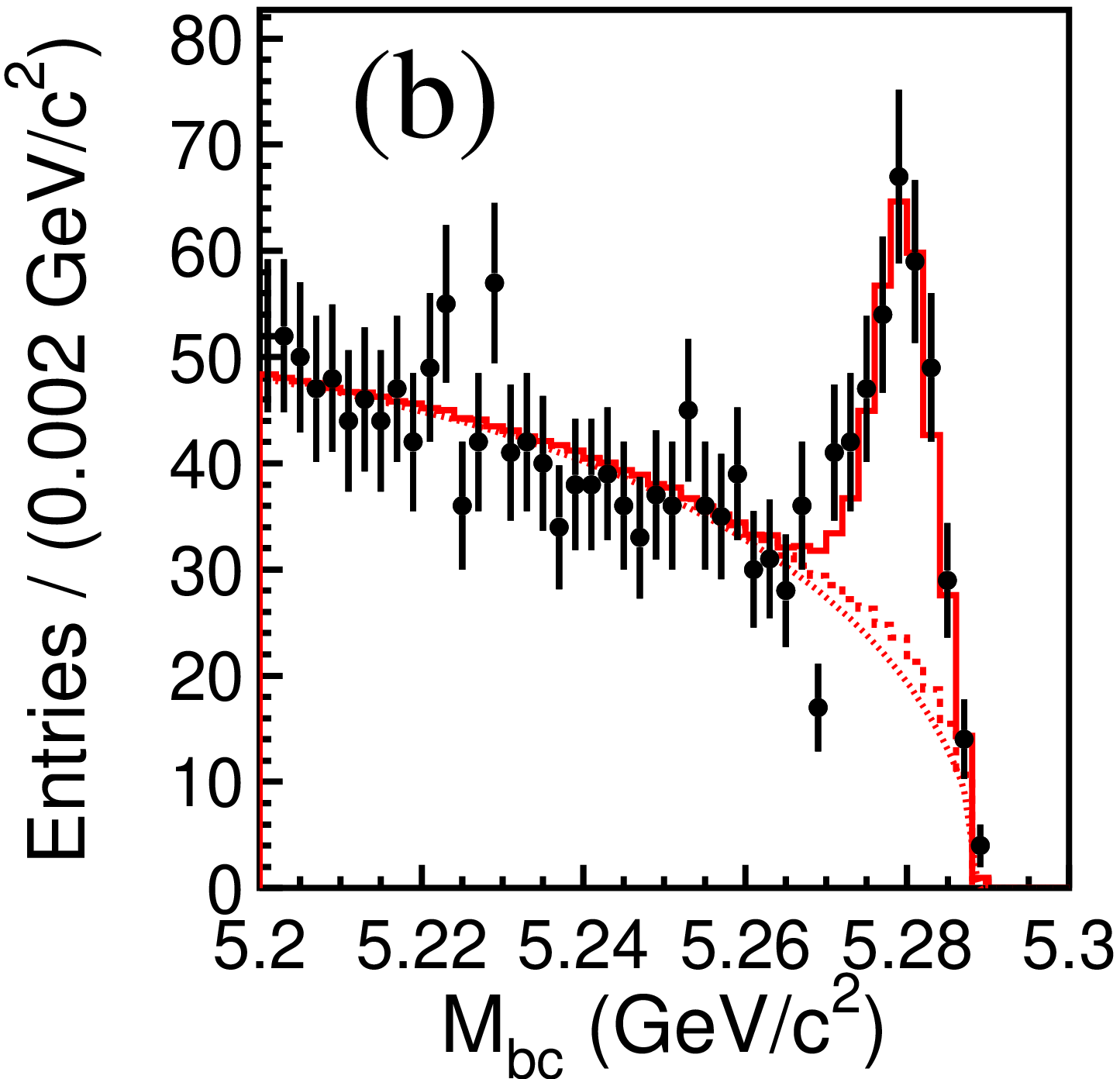}
\caption{
  (a) $\Delta E$ distribution within the $M_{\rm bc}$ signal slice
  and (b) $M_{\rm bc}$ distribution within the $\Delta E$ 
  signal slice for the
  whole $M_{K_S^0\pi^0}$ region. Points with error bars are data.
  The solid curves show the fit results. The dotted curves show the
  $q\overline{q}$ background contributions, 
  while the dashed curves show the sum
  of $q\overline{q}$ and $B\overline{B}$ background contributions.
} 
\label{fig:figure1}
\end{figure}
\par The $B$ candidates are identified using 
two kinematic variables: the energy difference 
$\Delta E \equiv E_B^{\rm cms} - E_{\rm beam}^{\rm cms}$ and the
beam-energy-constrained mass 
$M_{\rm bc} \equiv \sqrt{(E_{\rm beam}^{\rm cms})^2 - (p_B^{\rm cms})^2}$,
where $E_{\rm beam}^{\rm cms}$ is the beam energy in the cms, and 
$E_B^{\rm cms}$ and $p_B^{\rm cms}$ are the cms energy and momentum, 
respectively, of the reconstructed $B$ candidate.
The signal region in $\Delta E$ and $M_{\rm bc}$, which is used for the
measurements of $CP$-violating parameters,
is defined as $-0.2\;{\rm GeV} < \Delta E < 0.1\;\rm{GeV}$ and
$5.27 \;{\rm GeV/}c^2 < M_{\rm bc} < 5.29 \;{\rm GeV/}c^2$.
\par
\begin{figure}[h]
\centering
\includegraphics[width=40mm]{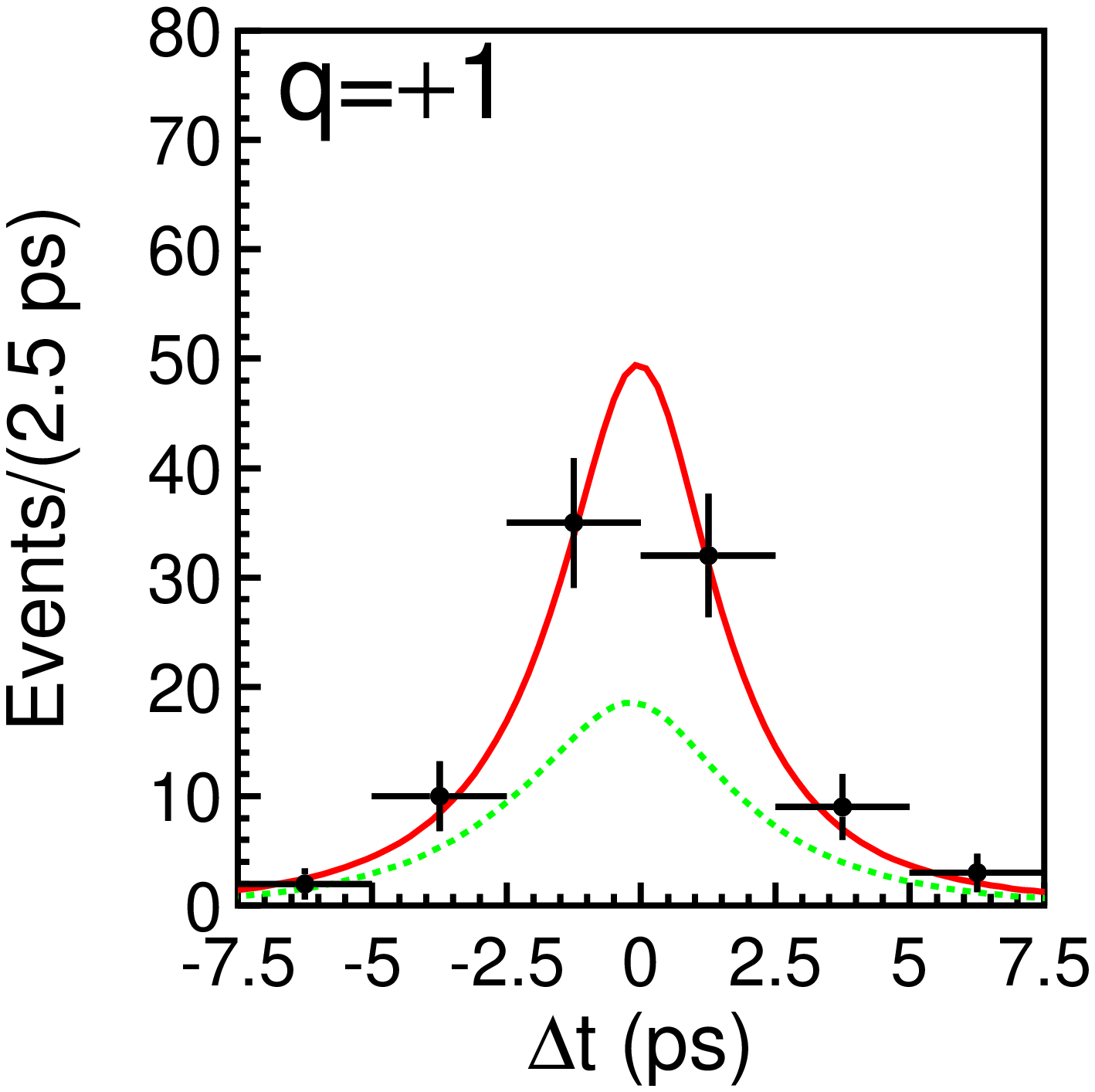}
\includegraphics[width=40mm]{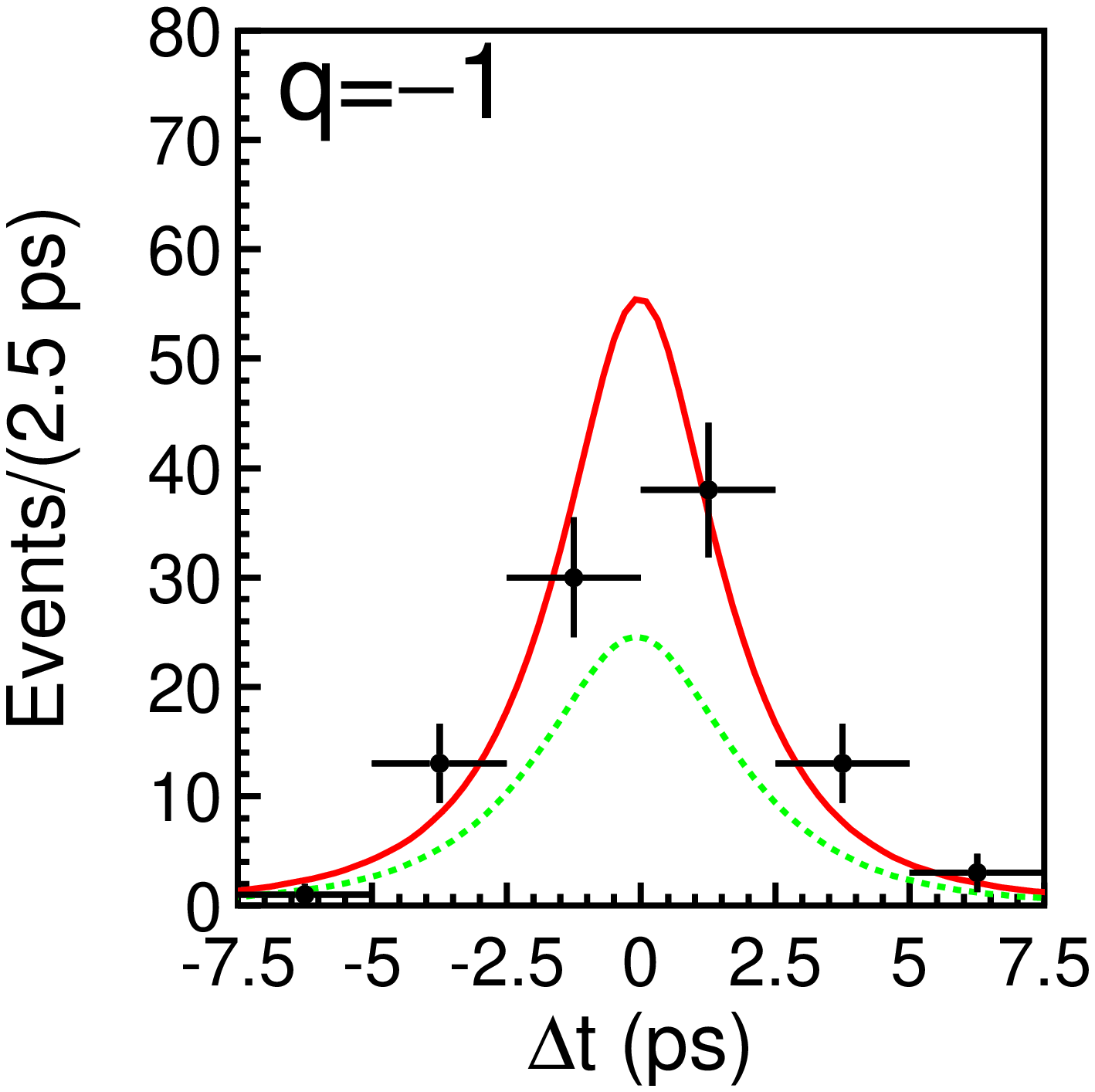}
\includegraphics[width=40mm]{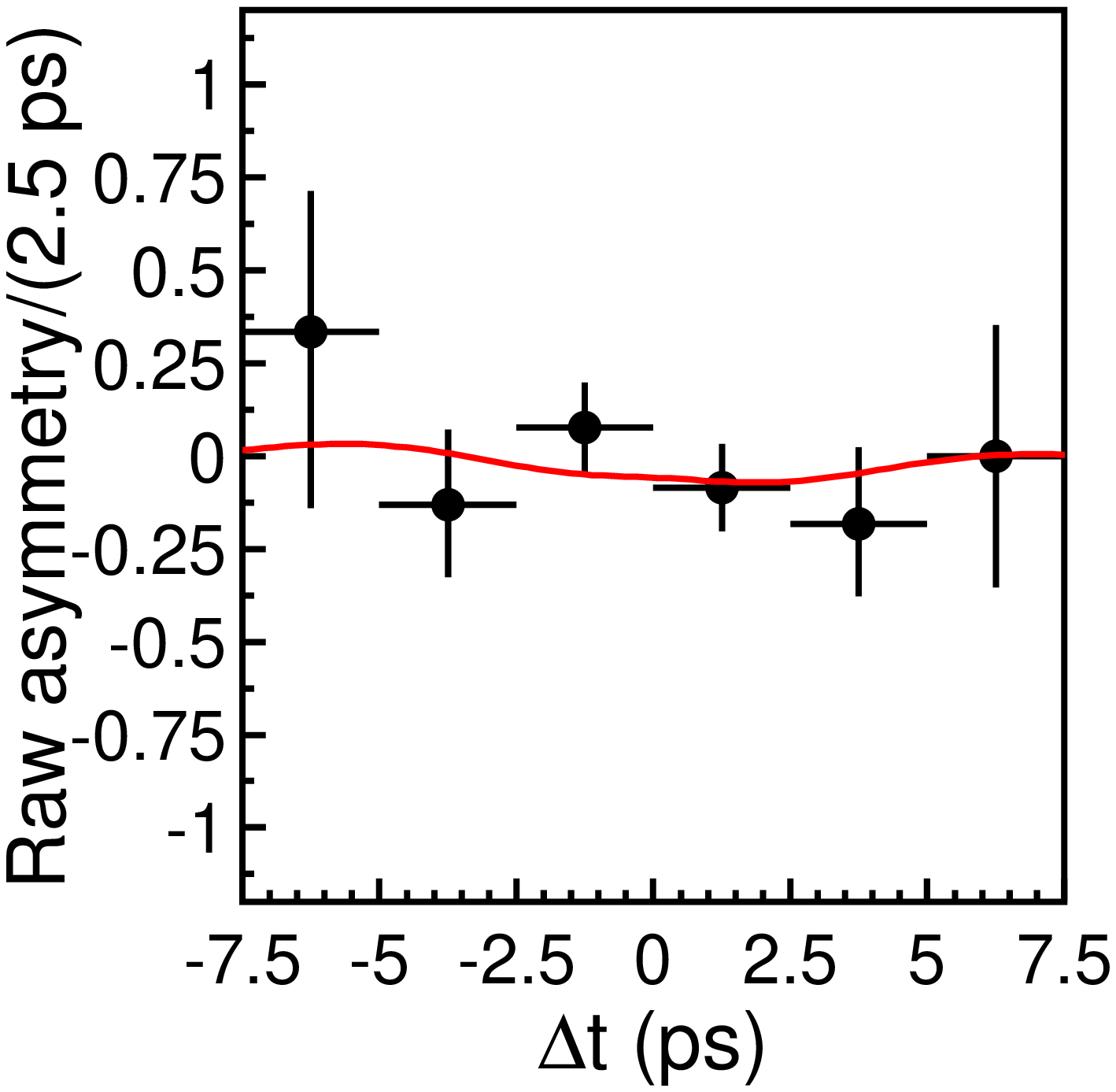}
\caption{
  (Top) Proper time distributions for $\btokpiog$
  for $q=+1$ (left) and $q=-1$ (right) with $0.5 < r \le 1.0$.
  The solid curve shows the total and dashed curve shows 
  the signal component.
  (Bottom) Asymmetry in each $\Delta t$ bin with $0.5 < r \le 1.0$.
  The solid curve shows the result of the UML fit.
}
\label{fig:figure2}
\end{figure}

 After all selections are 
applied, we obtain $4078$
candidates in the $\Delta E$-$M_{\rm bc}$ fit region,
of which $406$ are in the signal box. 
The signal yield is obtained from an 
UML fit to
the $\Delta E$-$M_{\rm bc}$ distribution as shown 
in Fig.~\ref{fig:figure1}.
Figure~\ref{fig:figure2}
shows the $\Delta t$ distributions of the events with $0.5 < r \le 1.0$
for $q=+1$ and $q=-1$ and the raw asymmetry.
The $CP$ violation parameters for $K_S\pi^0\gamma$ for the full
$K_S^0\pi^0$ invariant mass region
as well as for the mass region around $K^*(892)^0$ 
are summarized in Table~\ref{tab:table1}.
%
\section{Time-dependent Analysis of {\boldmath $\btokrhog$} }
The first measurement of time-dependent 
$CP$ asymmetry in $\btokrhog$ mode was performed by 
Belle using $657 \times 10^6$ 
$B\overline{B}$ pairs~\cite{lijin-ksrhogamma}.
The advantage of this mode is that the $B^0$ decay vertex
can be reconstructed from two charged pions from the $\rho^0$ decays, 
thus avoiding the complications and efficiency loss 
from $K_S^0$ vertexing. The expected $\mathcal{S}$ has opposite sign
to that of $\btokpiog$.
\par The signal is reconstructed in the decay $\btokrhog$ 
with $\rho^0 \to \pi^+\pi^-$ and $K_S^0 \to \pi^+\pi^-$.
The selection criteria for high energy prompt photon and 
neutral kaons are same
as those described in section~\ref{kspi0g},
except the photons are required 
to lie in the barrel region of the ECL and the 
$K_S^0$ invariant mass should be within 
$\pm 15\;{\rm MeV}/c^2$ of its nominal mass.
We also reconstruct the  $B^+\to K^+\pi^-\pi^+\gamma$ decay to
study the $K\pi\pi$ system and to serve as a control sample.
The $K^+\pi^-\pi^+$ and $K_S^0\pi^-\pi^+$
invariant masses are required to be less than 1.8 GeV/$c^2$.
The $B$ candidates are selected 
from the $K_S^0\pi^+\pi^- \gamma$ sample by requiring the
$\pi^+\pi^-$ invariant mass to lie in the $\rho^0$ region,
$0.6\,\mathrm{GeV}/c^2<m_{\pi\pi}<0.9\,\mathrm{GeV}/c^2$.
Since the $\rho^0$ is wide, other modes that are not self-conjugate, 
such as $K^{*+}\pi^- \gamma$ may also contribute.
Therefore, we first measure the effective $CP$-violating
parameters, $\mathcal{S}_{\rm{eff}}$ and
$\mathcal{A}_{\rm{eff}}$, using the final sample
and then convert them to the $CP$-violating parameters
of $B^0\to K_S^0 \rho^0\gamma$ using a dilution factor $\mathcal{D}$,
described in Ref.~\cite{lijin-ksrhogamma}.
\par
\begin{figure}[h]
\centering
\includegraphics[width=80mm]{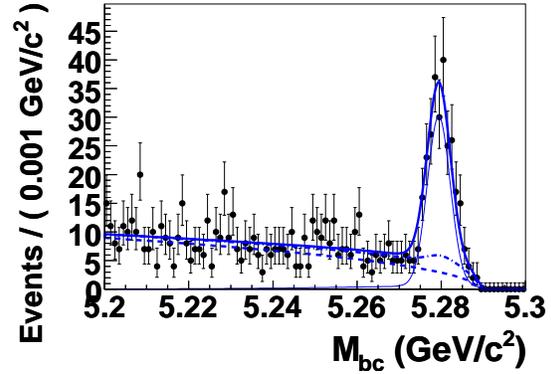}
\caption{
  $M_\mathrm{bc}$ distributions for 
  $B^0\to K_S^0\pi^+\pi^-\gamma$ events.
  Points with error bars are data.  The curves show the results
  from the $r$ dependent $M_\mathrm{bc}$ fit.
  The dashed and dash-dotted curves are the $q\bar q$ and all background.
  The thin curve is the total signal including SCF and the thick curve
  is the total PDF.} 
\label{fig:figure3}
\end{figure}

 The signal is extracted from an 
UML fit to the $M_\mathrm{bc}$ distribution 
as shown in Fig.~\ref{fig:figure3}.
The requirement $-0.1\,\mathrm{GeV}<\Delta E<0.08\,\mathrm{GeV}$
is applied.
We obtain 299 events in the signal 
$M_\mathrm{bc}$ region after vertexing.
Out of these we find a signal of $212\pm 17$ events 
with a fraction of 6.0\% self-cross-feed (SCF),
$53.4\pm 2.6$ continuum, along with 7.8 $K^*\gamma$,
21.5 other $X_s\gamma$, and 9.0 $B\bar B$ events.
\par
\begin{figure}[h]
\centering
\includegraphics[width=50mm]{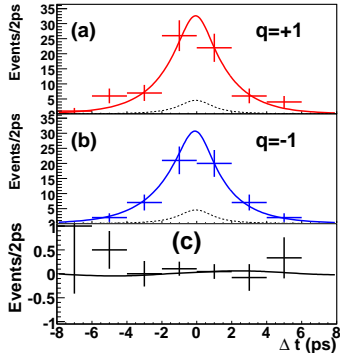}
\caption{
  Fit projections on the $\Delta t$ distributions with 
  (a) $q = +1$
  and (b) $q = -1$ for events with $r>0.5$. The solid curves are the
  fit while the dashed curves show the background contributions.
  The raw asymmetry as a function of $\Delta t$ is shown in (c)
  with a fit curve superimposed.} 
\label{fig:figure4}
\end{figure}

\begin{figure}[h]
\centering
\includegraphics[width=50mm]{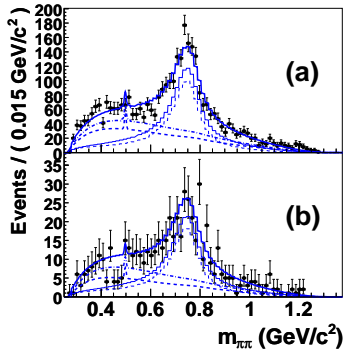}
\caption{
  $m_{\pi\pi}$ distributions for 
  (a) $B^+\to K^+\pi^-\pi^+\gamma$ and
  (b) $B^0\to K_S^0\pi^+\pi^-\gamma$ .
  The curves follow the convention in Fig.~\ref{fig:figure3}.
  The thin dashed curve is the correctly reconstructed 
  $B\to K_1(1270)\gamma$ signal.} 
\label{fig:figure5}
\end{figure}

 We obtain
$\mathcal{S}_\mathrm{eff} = 0.09\pm 0.27(\mathrm{stat.})
 ^{+0.04}_{-0.07}(\mathrm{syst.})$ 
and
$\mathcal{A}_\mathrm{eff} = 0.05\pm 0.18(\mathrm{stat.})
 \pm 0.06(\mathrm{syst.})$
from an 
UML fit to the 
observed $\Delta t$ distribution.
The parameter $\mathcal{S}_\mathrm{eff}$
is related to $\mathcal{S}$ for $K_S^0\rho^0\gamma$
with a dilution factor $\mathcal{D} 
\equiv \mathcal{S}_\mathrm{eff}/{\mathcal{S}_{K_S^0\rho^0\gamma}}$,
that depends on the $K^{*\pm}\pi^{\mp}$ components and 
allows for interference:
\begin{equation}
\label{eqn:s_d}
\mathcal{D} = 
 \frac{\int
[|F_A|^2+ 2\,{\rm Re}(F_A^*F_B) + F_B^*(\bar K)F_B(K) ]}
{\int \left[|F_A|^2+ 2\,{\rm Re}(F_A^*F_B) + |F_B|^2\right]},
\end{equation}
where $F_A,F_B$ are photon-helicity averaged amplitudes for
$B^0\to K_S^0\rho^0(\pi^+\pi^-)\gamma$ and 
$B^0\to K^{*\pm}(K_S^0\pi^\pm)\pi^\mp\gamma$, 
respectively.
The factors $F_B(\bar K)$, $F_B(K)$ distinguish between 
$K^{*-}\pi^+\gamma$ and
$K^{*+}\pi^-\gamma$.  The phase space integral is over 
the $\rho^0$ region.
We measure the dilution factor $\mathcal{D}$ 
to be $0.83^{+0.19}_{-0.03}$ from the charged mode 
$B^+\to K^+\pi^-\pi^+\gamma$
using a combination of various kaonic resonances with spin $\geq 1$
to model the $K\pi\pi$ system. 
By combining it with $\mathcal{S}_\mathrm{eff}$, we obtain
$\mathcal{S}_{K_S\rho^0\gamma} = 0.11\pm 0.33(\mathrm{stat.})
^{+0.05}_{-0.09}(\mathrm{syst.}) $. The fits to the observed $\Delta t$
distributions and the raw asymmetry are shown in the 
Fig.~\ref{fig:figure4}. 
The fit results are summarized in Table~\ref{tab:table1}.
Figure~\ref{fig:figure5} shows
the distributions for the $m_{\pi\pi}$ spectrum.
\begin{table}[h]
\caption{Results of the fits to the $\Delta t$ distributions. 
The first errors are statistical and the second errors are systematic.}
\begin{tabular}{lcc}
\hline
\hline
Decay Mode & $\mathcal{S}$ & $\mathcal{A}$ \\
  \hline
$B^0 \to K^*(892)^0 \gamma$ & $-0.32^{+0.36}_{-0.33}\pm0.05$ & $-0.20\pm0.24\pm0.05$ \\
$B^0 \to K_S \pi^0 \gamma$ & $-0.10\pm0.31\pm0.07$ & $-0.20\pm0.20\pm0.06$ \\
$B^0 \to K_S \rho^0 \gamma$ & $0.11\pm0.33^{+0.05}_{-0.09}$ & $0.05\pm0.18\pm0.06$ \\
\hline
\hline
\end{tabular}
\label{tab:table1}
\end{table}
\section{ First Observation of {\boldmath $\btopksg$} decay}
This decay mode has advantages similar to 
$\btokrhog$ in the search for new physics. 
Here the $B^0$ decay vertex can be reconstructed from the two
charged kaons from the $\phi$ decay.
The branching fractions for the charged $\phi K \gamma$ mode 
and an upper limit on the neutral mode have 
already been reported 
by the Belle~\cite{alex_prl} using 
$96 \times 10^6$ $B\overline{B}$ pairs.
We search for the neutral mode using the full data sample,
nearly eight times larger than was used in 
our previous measurement and report the first observation with a
significance of $5.4\,\sigma$.
\par The signal is reconstructed in the decay $\btopkpg$ 
and $\btopksg$, with 
$\phi \to K^+K^-$ and $K_S^0 \to \pi^+\pi^-$.
The invariant mass of the $\phi$ candidates is required to be within
$-0.01 < M_{K^+K^-}-m_{\phi} < +0.01 \;{\rm GeV}/c^2$,
where $m_{\phi}$ denotes the world-average $\phi$ mass~\cite{pdg}.
The selection criteria for the high energy prompt photon and 
neutral kaons are the same
as those described in section~\ref{kspi0g},
except the photons are required 
to lie in the barrel region and 
the invariant mass of the $\pi^+\pi^-$ combinations
has to be in the range 0.482 GeV/$c^2 < M_{\pi^+\pi^-} < 0.514$ GeV/$c^2$.
The $B$ candidates are selected with a requirement
$5.2 \;{\rm GeV/}c^2 < M_{\rm bc} < 5.3 \;{\rm GeV/}c^2$ and 
$-0.3 \;{\rm GeV} < \Delta E < 0.3 \;\rm{GeV}$.
We define the signal region as 
$5.27 \;{\rm GeV/}c^2 < M_{\rm bc} < 5.29 \;{\rm GeV/}c^2$ and 
$-0.08 \;{\rm GeV} < \Delta E < 0.05 \;\rm{GeV}$.
\par The dominant background is from the 
continuum process, which is 
suppressed by a requirement on likelihood ratio 
from event shape variables and 
the $B$ flight direction. In the $\btopksg$ mode,
some backgrounds from $b \to c$ decays, like
$D^0\pi^0$, $D^0\eta$ and $D^-\rho^+$
peak in the $M_{\rm bc}$ distribution. 
We remove these backgrounds by applying
a veto on the $\phi K_S^0$ invariant mass. 
The non-resonant background $B \to K^+ K^- K \gamma$, 
which peaks in the
$\Delta E$-$M_{\rm bc}$ signal region, is estimated 
using the $\phi$ mass sideband in data.
\par
\begin{figure*}[t]
\centering
\includegraphics[width=60mm]{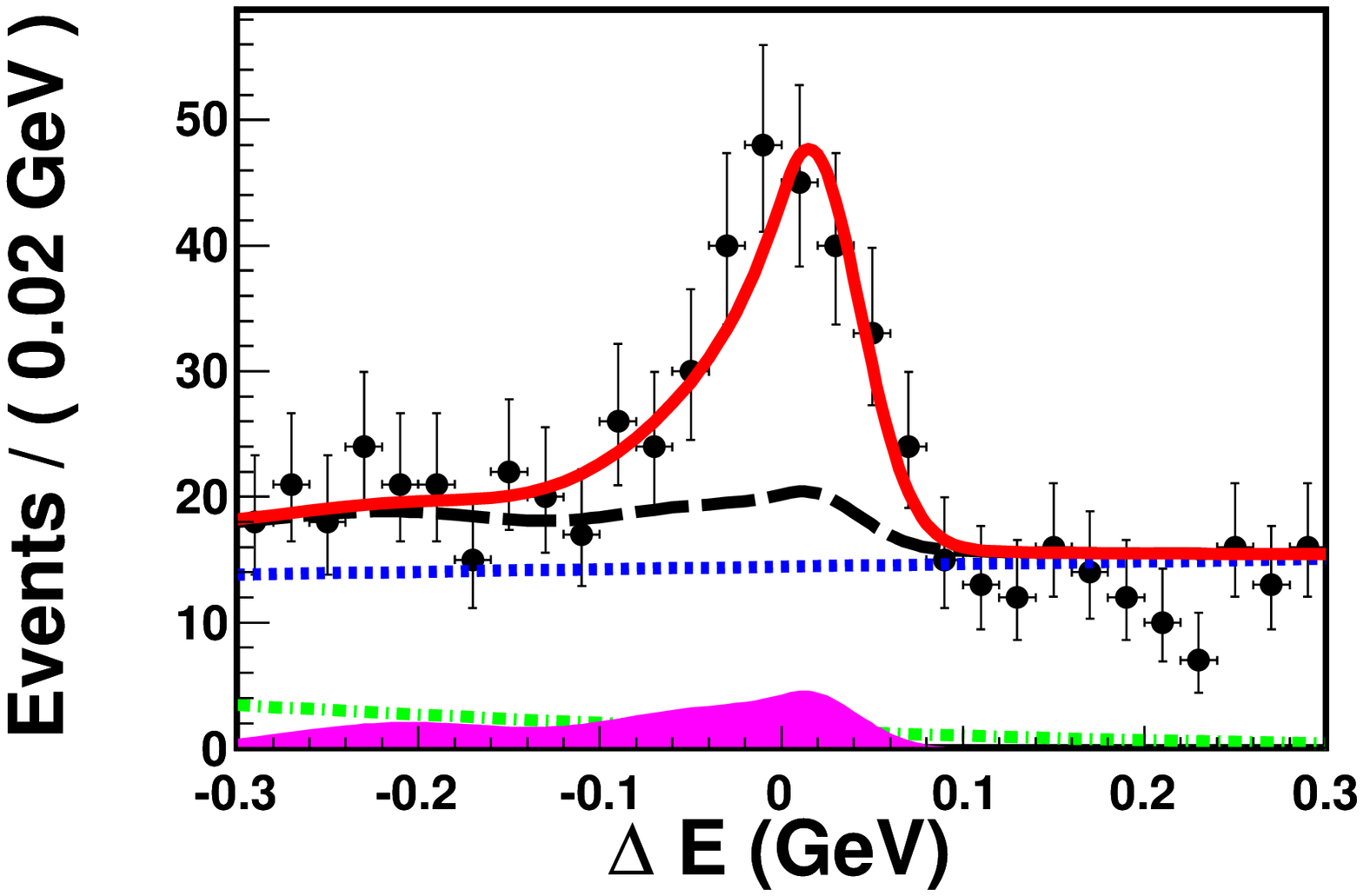}
\includegraphics[width=60mm]{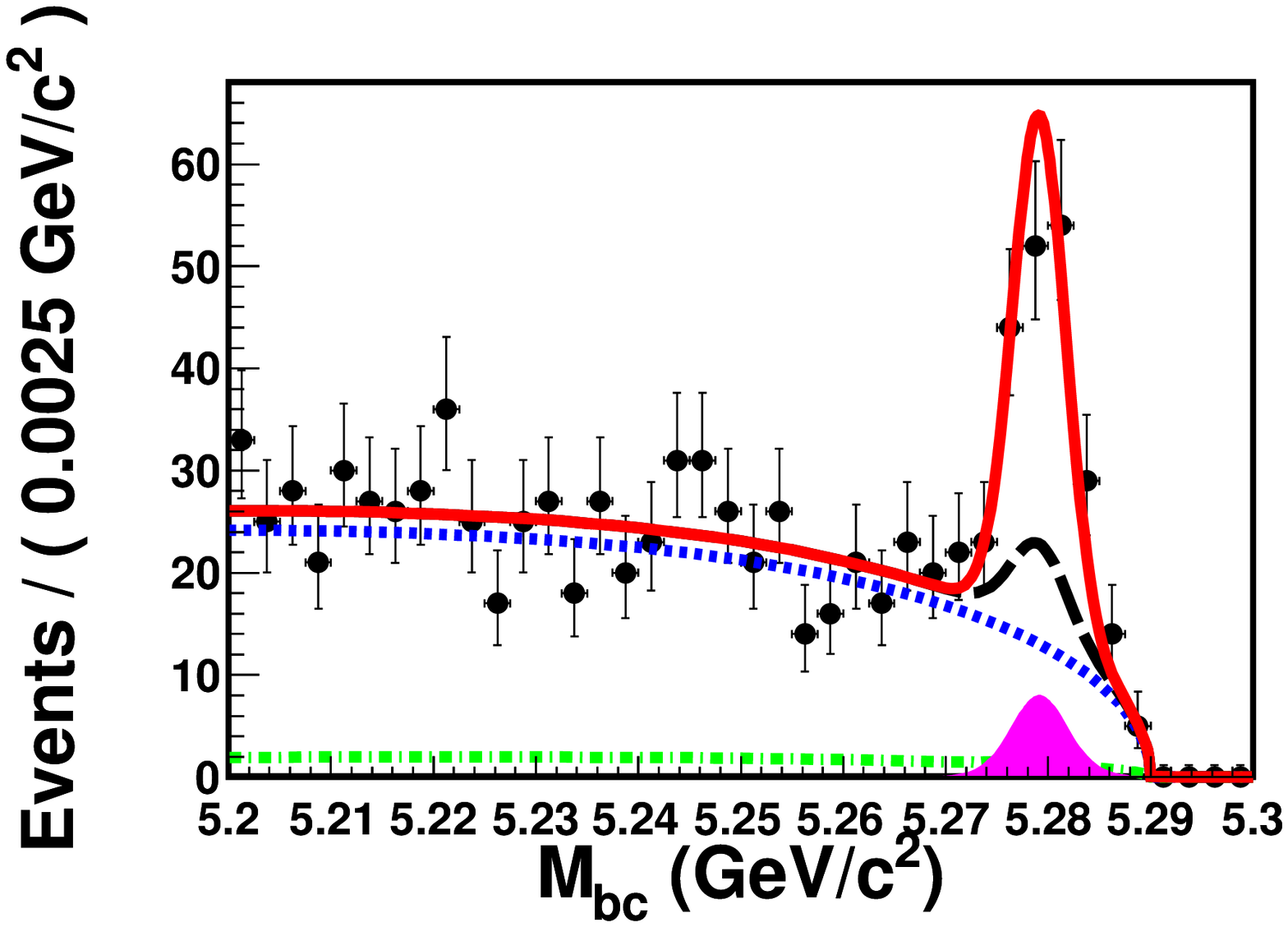}
\includegraphics[width=60mm]{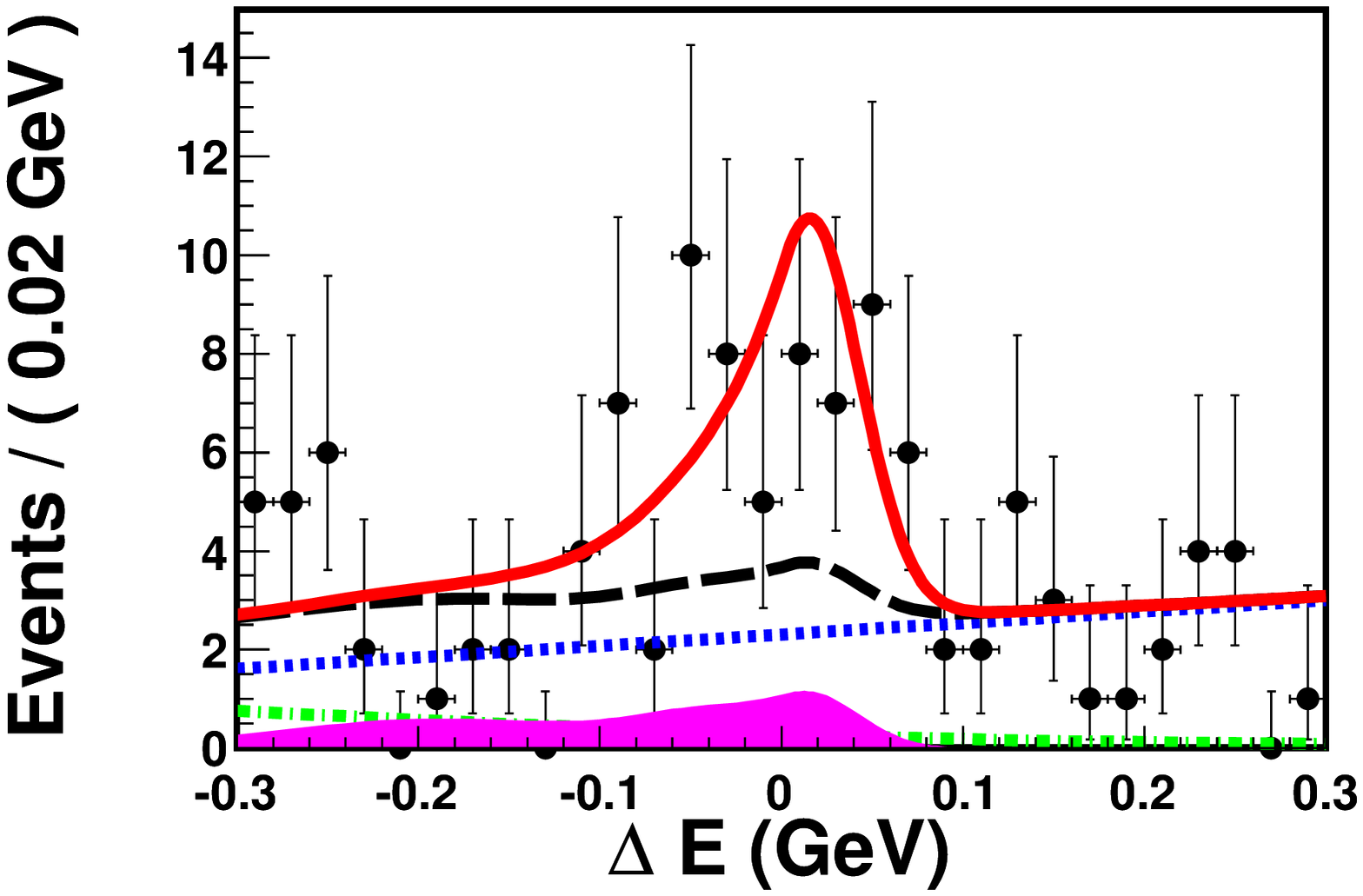}
\includegraphics[width=60mm]{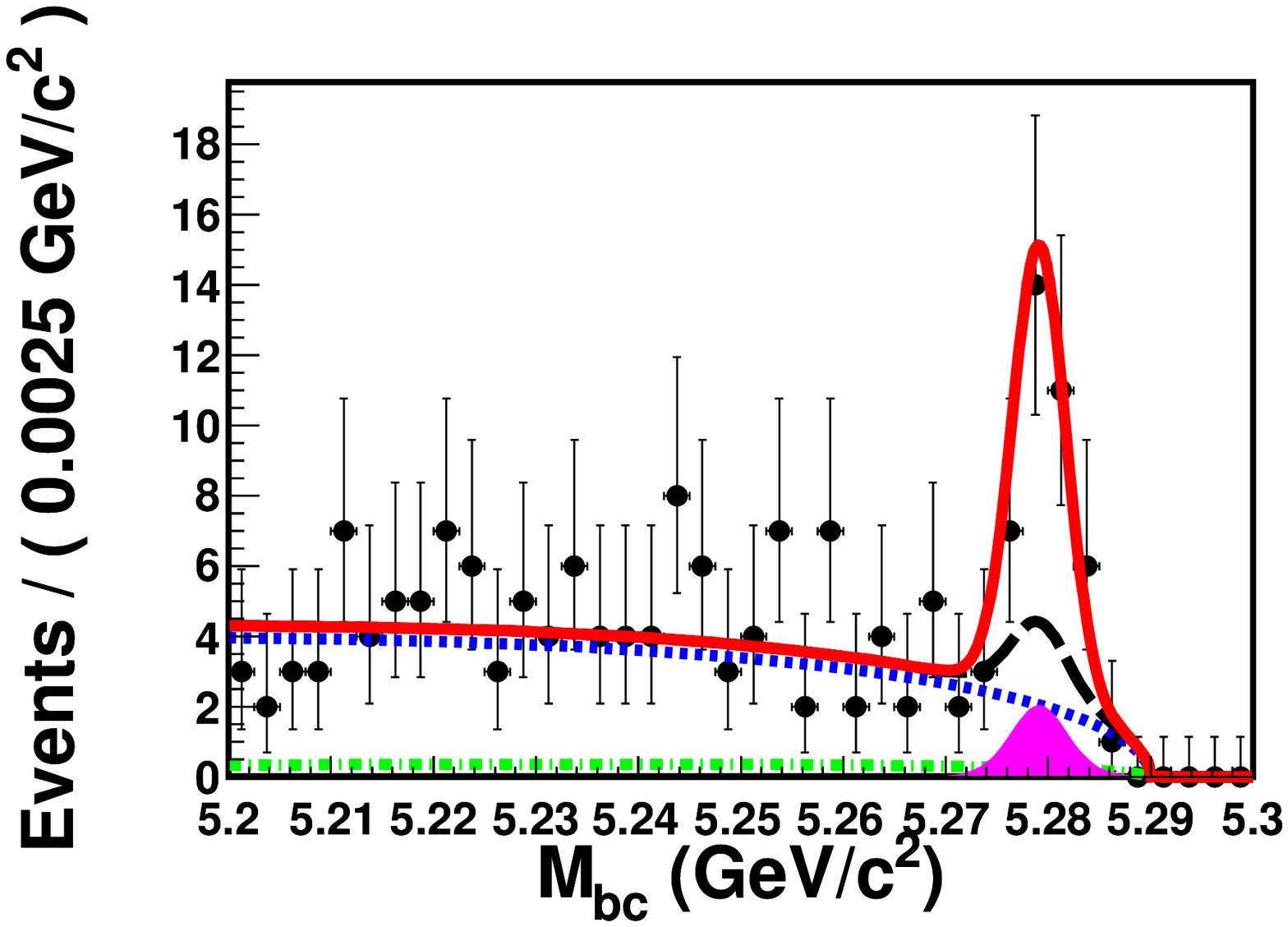}
\caption{The $\Delta E$ and $M_{\rm bc}$ projections 
for $\btopkpg$ (upper) and $\btopksg$ (lower). 
The points with error bars represent the data. The different curves 
show the total fit function (solid, red),   
total background function (long-dashed, black), 
continuum component (dotted, blue), 
the $b\to c$ component (dashed-dotted, green)
the non-resonant component as well as other charmless 
backgrounds (filled histogram, magenta).}
\label{fig:figure6}
\end{figure*}

 The signal yield is obtained from an extended
UML fit to the two-dimensional 
$\Delta E$-$M_{\rm bc}$ distribution. The projections of the 
fit results into $\Delta E$ and $M_{\rm bc}$ are shown in 
Fig.~\ref{fig:figure6} and the fit results are summarized in
Table~\ref{tab:table2}. The fit yields a signal of 
($136\pm17$) $\btopkpg$ and ($35\pm8$) $\btopksg$ candidates.
The signal in the charged mode has a significance of
$9.6\,\sigma$, whereas that for the neutral mode is $5.4\,\sigma$, 
including systematic uncertainties.
\begin{table}[h]
\caption{The signal yields ($Y$), corrected efficiencies ($\epsilon$), 
branching fractions ($\mathcal{B}$) and significances ($\mathcal{S}$)
for the $\btopkpg$ and $\btopkog$ decay modes.}
\begin{tabular}{lcccc}
\hline
\hline
Mode & $Y$ & $\epsilon$ ($\%$) & $\mathcal{B}$ ($10^{-6}$) & $\mathcal{S}$ ($\sigma$) \\
  \hline
$\phi K^+ \gamma$ & $136\pm17$ & $15.3\pm0.1$ & $2.34\pm0.29\pm0.23$ & $9.6$ \\
$\phi K^0 \gamma$ & $35\pm8$ & $10.0\pm0.1$ & $2.66\pm0.60\pm0.32$ & $5.4$ \\
\hline
\hline
\end{tabular}
\label{tab:table2}
\end{table}
%
\par
\begin{figure}[h]
\centering
\includegraphics[width=57mm]{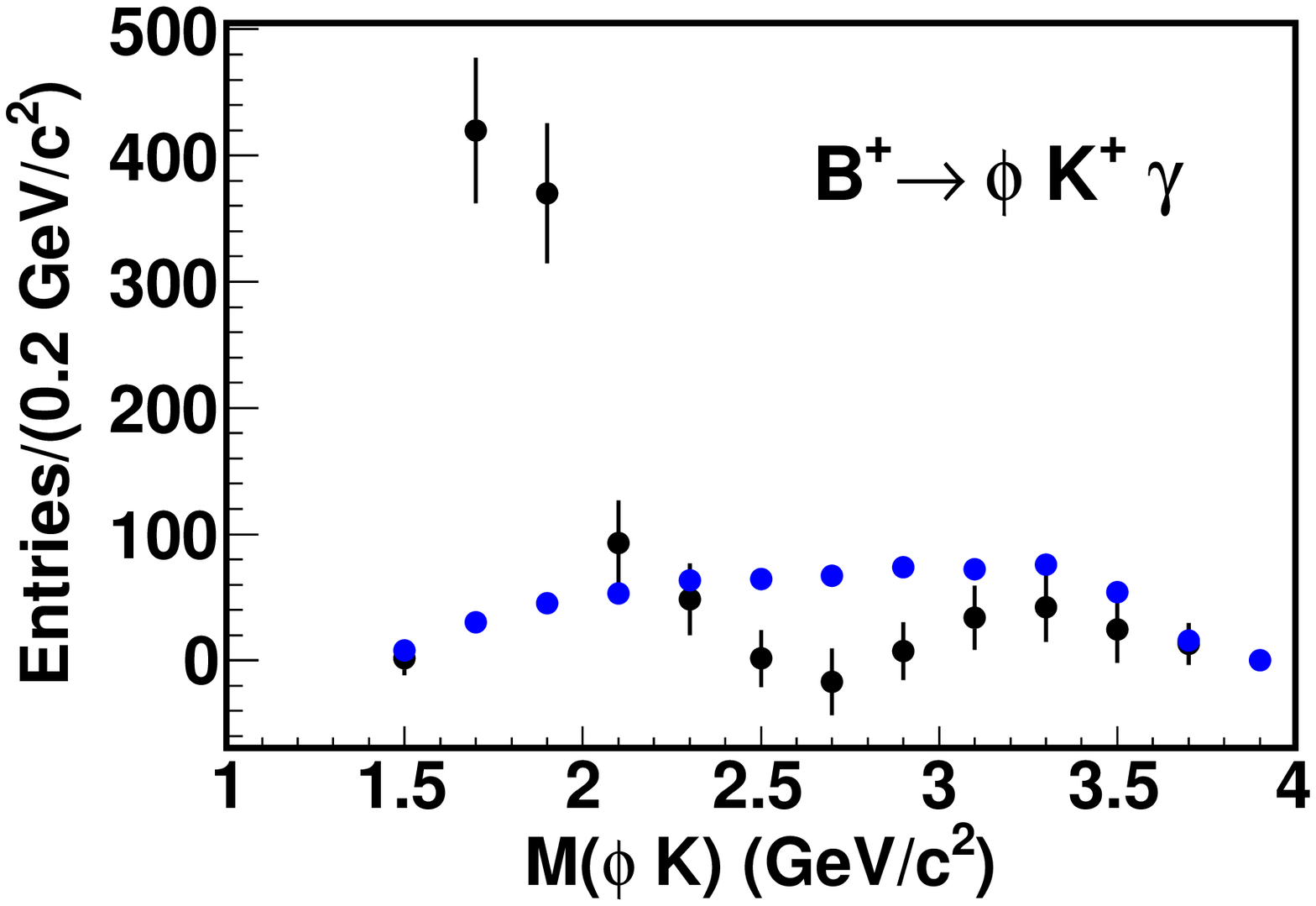}
\includegraphics[width=57mm]{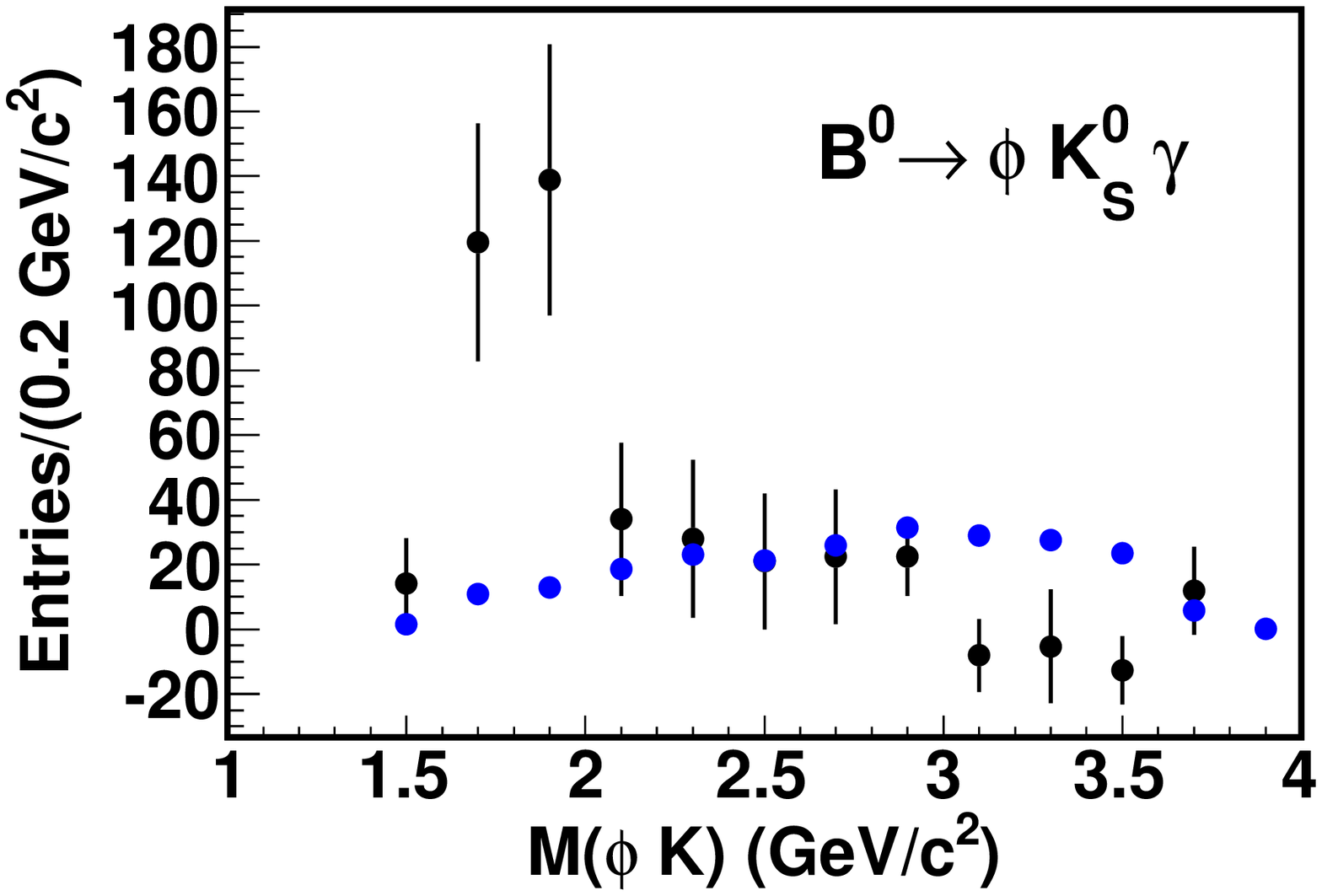}
\caption{The background-subtracted and efficiency-corrected 
$\phi K$ mass distributions for 
$\btopkpg$ (upper) and $\btopksg$ (lower). 
The points with error bars represent the data. The
yield in each bin is obtained by the fitting procedure 
described in the text. The three-body phase-space model from 
the MC simulation is shown by the circles (blue).}
\label{fig:figure7}
\end{figure}

 We also search for a possible contribution from 
kaonic resonances decaying to $\phi K$. 
To unfold the $M_{\phi K}$ distribution, 
we subtract all possible backgrounds and correct the
$\phi K$ invariant mass for the efficiency. 
The background-subtracted and efficiency-corrected 
$M_{\phi K}$
distributions are shown in Fig.~\ref{fig:figure7}.
Nearly $72\%$ of the signal events are concentrated
in the low-mass region ($1.5 < M_{\phi K} <2.0$ GeV/$c^2$). 
It is clear that the observed 
$\phi K$ mass spectrum differs significantly from that expected in a 
three-body phase-space decay.
\section{Summary}
\begin{figure}[h]
\centering
\includegraphics[width=80mm]{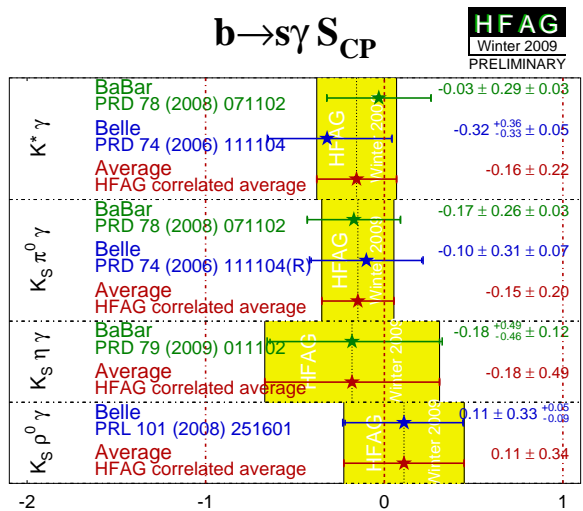}
\caption{Summary of Belle and BaBar measurements of $b\to s \gamma$ $S_{\rm CP}$.}
\label{fig:figure8}
\end{figure}
In summary, we report the first observation of radiative 
$\btopksg$ decays in Belle using a data sample of 
$772 \times 10^6$ $B\overline{B}$ pairs. The observed signal yield is
($35\pm8$) with a significance of $5.4\,\sigma$ 
including systematic uncertainties.
We also precisely measure the $\btopkpg$ branching fraction
with a significance of $9.6\,\sigma$. 
The signal events are mostly concentrated at low $\phi K$ mass,
which is similar to a two-body radiative decay.
We report the measurements of time-dependent 
$CP$ violation parameters in $\btokpiog$ and $\btokrhog$
decays, using $535 \times 10^6$ and 
$657 \times 10^6$ $B\overline{B}$ pairs, respectively.
The Heavy Flavor Averaging Group (HFAG)~\cite{hfag} 
summary of the measurements of the parameter
$S$ in $b\to s \gamma$ modes by the Belle and BaBar is shown in 
Figure~\ref{fig:figure8}.
With the present statistics, 
these measurements are consistent with the 
standard model predictions and
there is no indication of New Physics from right-handed 
currents in radiative $B$ decays.
The neutral $\btopksg$ mode has enough statistics 
for a future measurement of time-dependent $CP$ violation. 
More luminosity is necessary for a precise test of the SM.
\begin{acknowledgments}
We thank the KEKB group for excellent operation of the
accelerator, the KEK cryogenics group for efficient solenoid
operations, and the KEK computer group and
the NII for valuable computing and SINET3 network support.  
We acknowledge support from MEXT, JSPS and Nagoya's TLPRC (Japan);
ARC and DIISR (Australia); NSFC (China); 
DST (India); MEST, KOSEF, KRF (Korea); MNiSW (Poland); 
MES and RFAAE (Russia); ARRS (Slovenia); SNSF (Switzerland); 
NSC and MOE (Taiwan); and DOE (USA).
\end{acknowledgments}
\bigskip 

\begin{thebibliography}{9}   

\bibitem{ags1}
D.~Atwood, M.~Gronau and A.~Soni, Phys. Rev. Lett. {\bf 79}, 185 (1997).
\bibitem{ags2}
D.~Atwood, T.~Gershon, M.~Hazumi and A.~Soni, Phys. Rev. D {\bf 71}, 076003 (2005).
\bibitem{conj}
Throughout this paper, the inclusion of the charge-conjugate 
decay mode is implied unless otherwise stated.
\bibitem{Belle}
A.~Abashian {\it et al.} (Belle Collaboration), 
Nucl. Instrum. Methods Phys. Res., Sect. A {\bf 479}, 117 (2002).
\bibitem{kekb}
S.~Kurokawa and E.~Kikutani, Nucl. Instrum. Methods Phys. Res., Sect. A 
{\bf 499}, 1 (2003), and other papers included in this volume.
\bibitem{tag}
H.~Kakuno {\it et al.}, 
Nucl. Instrum. Methods Phys. Res., Sect. A 
{\bf 533}, 516 (2004).
\bibitem{belle_cc}  
K.~Abe {\it et al.} (Belle Collaboration), 
Phys. Rev. D {\bf 71}, 072003 (2005).
\bibitem{belle_b2s}
K-F.~Chen {\it et al.} (Belle Collaboration), Phys. Rev. D {\bf 72},
012004 (2005).
\bibitem{ushiroda-kspi0gamma}
Y.~Ushiroda {\it et al.} (Belle Collaboration), Phys. Rev. D {\bf 74}, 111104(R) (2006).
\bibitem{pi0etaveto}
P.~Koppenburg {\it et al.} (Belle Collaboration), Phys. Rev. Lett. {\bf 93}, 061803 (2004).
\bibitem{lijin-ksrhogamma}
J.~Li {\it et al.} (Belle Collaboration), Phys. Rev. Lett. {\bf 101}, 251601 (2008).
\bibitem{alex_prl}
A.~Drutskoy {\it et al.} (Belle Collaboration), Phys. Rev. Lett. {\bf 92}, 051801 (2004).
\bibitem{pdg}
C.~Amsler, {\it et al.}, Physics Letters {\bf B 667}, 1 (2008).
\bibitem{hfag} 
Heavy Flavor Averaging Group, winter 2009 update.
Check their webpage for updated results: http://www.slac.stanford.edu/xorg/hfag/.



\end{thebibliography}

\end{document}